\documentclass[aps, prx, 10pt,american,twocolumn,nofootinbib,groupedaddress,floatfix, superscriptaddress,amsmath,amssymb,longbibliography,citeautoscript]{revtex4-2}

\usepackage[T1]{fontenc}
\usepackage{newtxtext}
\usepackage{newtxmath}

\usepackage{amsmath,tikz,calc}
\usepackage{mathtools}
\usepackage{bm}
\usepackage{xfrac}
\usepackage{braket}

\usepackage{color}
\usepackage{xcolor}
\usepackage[colorlinks,
            citecolor=blue,
            linkcolor=blue,
            urlcolor=blue]{hyperref}

\usepackage{graphicx}
\usepackage{subfigure}
\usepackage[all]{hypcap} % Fix reference to figures
\usepackage{lipsum}% http://ctan.org/pkg/lipsum

\usepackage{layout}
\usepackage{booktabs}
\usepackage{siunitx}

\usepackage{graphicx}% Include figure files
\usepackage[utf8]{inputenc}
\usepackage{hyperref}% add hypertext capabilities
\usepackage{color}
\usepackage{physics}

\definecolor{BLACK}{gray}{0}
\definecolor{WHITE}{gray}{1}
\definecolor{RED}{rgb}{1,0,0}
\definecolor{GREEN}{rgb}{0,1,0}
\definecolor{BLUE}{rgb}{0,0,1}
\definecolor{CYAN}{cmyk}{1,0,0,0}
\definecolor{MAGENTA}{cmyk}{0,1,0,0}
\definecolor{YELLOW}{cmyk}{0,0,1,0}
\definecolor{armygreen}{rgb}{0.55, 0.73, 0.0}

\newcommand{\ejd}[1]{{\color{BLACK} #1}}

\begin{document}

\title{A tale of two localizations: coexistence of flat bands and Anderson localization in a photonics-inspired amorphous system}

\author{Elizabeth J. Dresselhaus}
\thanks{These two authors contributed equally.}
\affiliation{Department of Physics, University of California, Berkeley, California 94720, USA}

\author{Alexander Avdoshkin}
\thanks{These two authors contributed equally.}
\affiliation{Department of Physics, University of California, Berkeley, California 94720, USA}
\affiliation{Department of Physics, Massachusetts Institute of Technology, Cambridge, MA 02139, USA}

\author{Zhetao Jia}
\affiliation{Department of Electrical Engineering and Computer Sciences, University of California, Berkeley, California 94720, USA}

\author{Matteo Secl\'{i}}
\affiliation{Department of Electrical Engineering and Computer Sciences, University of California, Berkeley, California 94720, USA}
\affiliation{École Polytechnique Fédérale de Lausanne (EPFL), 1015 Lausanne, Switzerland}

\author{Boubacar Kant\'{e}}
\affiliation{Department of Electrical Engineering and Computer Sciences, University of California, Berkeley, California 94720, USA}
\affiliation{Materials Sciences Division, Lawrence Berkeley National Laboratory, 1 Cyclotron Road, Berkeley, California 94720, USA}

\author{Joel E. Moore}
\affiliation{Department of Physics, University of California, Berkeley, California 94720, USA}
\affiliation{Materials Sciences Division, Lawrence Berkeley National Laboratory, 1 Cyclotron Road, Berkeley, California 94720, USA}

\date{\today}

\begin{abstract}

Emerging experimental platforms use amorphousness, a constrained form of disorder, to tailor meta-material properties. We study localization under this type of disorder in a family of 2D models generalizing recent experiments on photonic systems. \ejd{Models in this family reside on amorphous analogs of kagom\'{e} lattices with fixed coordination number, vary by a tunable synthetic field, and remarkabaly, permit exact results. We observe two kinds of localization that emerge in these models: Anderson localization by amorphous disorder}, and the existence of compact, macroscopically degenerate localized states as in many crystalline flat bands. The flat-band-like degeneracy innate to kagom\'{e} lattices survives under amorphousness without on-site disorder. This phenomenon arises from the cooperation between the structure of the compact localized states and the geometry of the amorphous graph. \ejd{More surprisingly}, for particular values of the field, such states emerge in the amorphous system that were not present on the kagom\'{e} lattice in the same field. \ejd{Outside the flat band, constrained amorphous graph geometry necessitates the existence of a fully delocalized state, near which we observe evidence of a localization-delocalization transition. Our platform serves as a  demonstration of how the qualitative behavior of a disordered system can be tuned at fixed graph topology and lead to localization phenomena unique to amorphous systems that are not observed in their generically disordered counterparts.}
\end{abstract}

\maketitle

\begin{section}{Introduction}
Amorphous materials are defined by exhibiting local order in the absence of long-range order. They have traditionally been preferred in optical applications due to their isotropic properties and robustness, while still maintaining the necessary spectral gap \cite{fox2010optical}. More recently, engineered  amorphousness has also been used in photonic meta-materials for similar reasons \cite{florescu2009designer}. Conversely, in electronics, crystalline structures are favored because of their higher electron mobility. This difference fundamentally arises from the distinct spectral properties of amorphous versus periodic systems that share similar local structure. The primary variation lies in the eigenmodes, whereas the permitted energy levels are relatively unaffected. 

In this work, we study the structure of amorphous spectra and corresponding wavefunctions to understand how amorphous systems behave in ways that are not typical for generic disordered systems. The absence of translational invariance in any disordered system poses a challenge: study of crystalline solids is to a high degree aided by Bloch's theorem which explains both the presence of band gaps and the extended nature of the wave functions. Solid-state physics has traditionally focused on perturbative disorder of a general form (typically, random on-site potential or hopping terms), but with small amplitude (weak disorder) \cite{girvin2019modern}. Amorphous disorder, in contrast, is an example of strong disorder of a very special form that is non-perturbative. It can be thought of as a discrete hopping disorder that ``rewires" the underlying connectivity graph of a system, but preserves the hopping amplitude. This property along with the absence of periodicity makes theoretical studies of amorphousness challenging.

We focus on an amorphous version of the kagom\'{e} lattice with a spatially uniform hopping phase and corresponding flux around closed loops of the underlying graph, analogous to a synthetic magnetic field. This model is realizable in resonator-array photonic devices \citep{hafezi2011} and has recently been studied by the present authors in this context \citep{zhetao-amorphousnl}. Additionally, it is structurally similar to the commonly studied models giving an approximate description of solids such as amorphous silicon or silicon dioxide \cite{catlow2012defects}.  The model preserves local order: it is defined on an hourglass graph, see Fig. \ref{fig:kagomization}, i.e., one composed of corner-sharing triangles.

%Quite remarkably, amorphous systems can host exotic phases of matter known to exist in crystalline systems. Recent experimental \cite{paul-amorphous-BiSe} and theoretical \citep{shenoy-agarwala, Marsal2020} research has demonstrated the persistence of topological phases in amorphous systems and quantum spin liquids have been predicted in amorphous systems \cite{cassella-amorphous-qsl}. 

Rigorous results for amorphous systems are scarce, with the notable exception of the demonstrated persistence of a spectral gap in amorphous tetrahedral bonded semiconductors despite disorder \cite{wearie, wearie-thorpe}. The robustness of the gap to amorphous disorder is markedly different from the case of random on-site disorder, where sharp crystalline bands broaden with increasingly strong disorder until the material eventually becomes a featureless insulator. In our work, we demonstrate and prove new exact results about the persistence of  macroscopically degenerate states coming from compactly localized states in the presence of amorphous disorder. In a slight misappropriation of language, we refer to these persistent highly degenerate energy levels as ``flat bands'' even though there is no Brillouin zone; they are flat bands as a function of external fluxes through a torus~\cite{niuthouless}. For general reviews of results on flat bands, see \cite{leykam2018artificial,photonicflatbands,danieli2024flat, rhim2021singular, neves2024crystal, mallick2022anti, flat_sharp_review}. Furthermore, quite remarkably, we show that flat bands exist in amorphous systems that have no analog in their crystalline counterparts. Flat bands are conducive to the formation of correlated states \cite{checkelsky2024flat}, such as ferromagentism \cite{mielke1991ferromagnetism}, superconductivity \cite{flat-band-review-superfluidity}, quantum spin liquids \cite{flat-band-review-qsl} and the fractional quantum hall effect \cite{flat-band-review-fqh}, and our model offers an interesting domain for such exploration as well as novel thermodynamic and transport features that accompany a spike in the density of states.

There are two standard pathways to the formation of flat bands \cite{rhim2021singular}: compact localized states due to destructive interference of hopping paths on a geometrically frustrated lattice, such as the kagom\'{e} \cite{chen-decoding-flatbands} or pyrochlore lattice \cite{pyrochlore}, and the Landau-level-like mechanism leading to topologically nontrivial flat bands\cite{landau_levels}. In the context of this work, we only encounter flat bands with compactly localized states (the first type). Flat bands are generically destroyed by on-site and other traditionally studied forms of disorder, although the states can stay localized \cite{leykam-2017}. Quite remarkably, it was noticed that disorder that preserves local connectivity \cite{sutherland1986localization, kohmoto1986electronic, schirmann2024physical} as well other types of specially engineered breaking of periodicity \cite{liu2022unconventional,ramachandran2017chiral, marsal2023obstructed, mallick2022anti} can leave flat bands and compact localized states therein intact. We observe similar phenomena for a general class of experimentally relevant amorphous models with partial or full preservation of the flat band degeneracy and non-trivial modification of the geometry of localized states.

Compact localized states are not the only localized states we observe in these models. Localization also originates from destructive interference of eigenmodes across randomized scattering paths. This phenomenon, known as Anderson localization \citep{anderson-localization}, has previously been observed for a class of our model \citep{zhetao-amorphousnl}. In the case of on-site disorder, a comprehensive paradigm has been established through application of thoroughly developed analytic tools \citep{efetov1999supersymmetry} and tests in numerical simulations \citep{EversMirlin:review}. In amorphous systems the fate of Anderson localization has been considered \citep{thesis-martin, Marsal2020, martin-multifracal, grim-localization-geometric-disorder, logan-wolynes} but a systematic study has not yet been undertaken. We begin to address this issue in this work by considering a model of amorphousness relevant to experiments in photonics. In particular, we find that, in some symmetry classes, for our model even the qualitative features of localization seem distinct from the standard paradigm.
 
The rest of the work is organized as follows. In Sec. \ref{sec:model} we introduce the model, defined on a kagom\'{e} lattice, and explain how we construct amorphous structures that are locally identical to kagom\'{e} lattices. In Sec. \ref{sec:flat} we explain why flat bands exist in some members of this family of model Hamiltonians and show how not only do these flat bands persist in the presence of amorphous disorder, there are in fact model Hamiltonians on amorphous structures that host flat bands which are absent in their periodic counterparts. We then turn to examining localization in the dispersive bands of our model. In Sec. \ref{sec:ALoc} we introduce a method to compute the effective localization length of eigenstates and adapt the method of energy level spacing distribution analysis to our amorphous systems. In Sec. \ref{sec:class_D} and \ref{sec:classA} we compare the localization properties of model systems tuned to be in two different symmetry classes. \ejd{Finally, in Sec. \ref{sec:ldl}, we discuss our observation of a localization-delocalization transition unique to amorphous systems.% with broken particle-hole symmetry.}
We explain how this behavior arises because of properties beyond the conventional symmetry classification of localization in disordered systems, chiefly the coexistence of randomness with constrained local coordination in amorphous systems.}

%We discuss differences and challenges that \ejd{these results raise} to

\end{section}

\section{Model}\label{sec:model}

The model we consider is motivated by experiments in photonic metamaterials \cite{hafezi2011}. The experimental platform consists of a $2D$ array of photonic resonators, where nearest neighbors are coupled to each other through two sets of waveguides. When the arm lengths of the waveguides within each pair differ, photon dynamics in the system are described by a magnetic tight-binding Hamiltonian \citep{hafezi2011} (in first quantization):

\begin{equation}
H_0 = \sum_{\Delta_i}e^{i\phi}(|i_1\rangle \langle i_2| + | i_2 \rangle\langle i_3| + |i_3 \rangle \langle i_1 |) + h. c.,
\label{eq: ham}
\end{equation}

where $\Delta_i$ is the $i$th triangle of the kagom\'{e} lattice. Sites are ordered $i_1, i_2, i_3$ as a counterclockwise path around each triangle. The effective magnetic field globally averages to zero but does not vanish locally for $\phi \neq 0$. 

 We can access three different Altland-Zirnbauer (AZ) symmetry classes \cite{altlandzirnbauer} simply by tuning the hopping phase $\phi$. At generic $\phi$ the system possesses neither chiral nor particle-hole nor time reversal symmetry (AZ symmetry class A). Time reversal symmetry is preserved only when $\phi = 0, \pi$ (AZ symmetry class AI). When $\phi = \pi/2$, the Hamiltonian has particle hole symmetry under complex conjugation $H_0^{*} = -H_0$ and falls into AZ class D. 

We will first consider the model of Eq. \eqref{eq: ham} on a (periodic) kagom\'{e} lattice. In the amorphous case, we define the model on graphs created by kagomization (explained in the next section) and refer to the analogous entities as amorphous systems. We note that this model is experimentally relevant at finite system size - photonic systems in experiments can have size $L \sim 100$, where $L$ denotes the number of resonators per row.

\subsection{Amorphous Model}

We adapt the amorphous graphs we used in \citep{zhetao-amorphousnl}: these graphs resemble kagom\'{e} lattices locally in that every vertex has coordination number $z = 4$ and is the center of a pair of corner-sharing triangles, however the graphs lack long range order and contain faces
other than triangles and hexagons. Some experimental realizations of photonics systems, including the experiments in our previous work, require equal bond length between all connected sites. We relax this restriction in order to have a simple process of generating graphs that will be easily reproducible.

\begin{figure}[t]
\centering
\includegraphics[width = 0.45\textwidth]{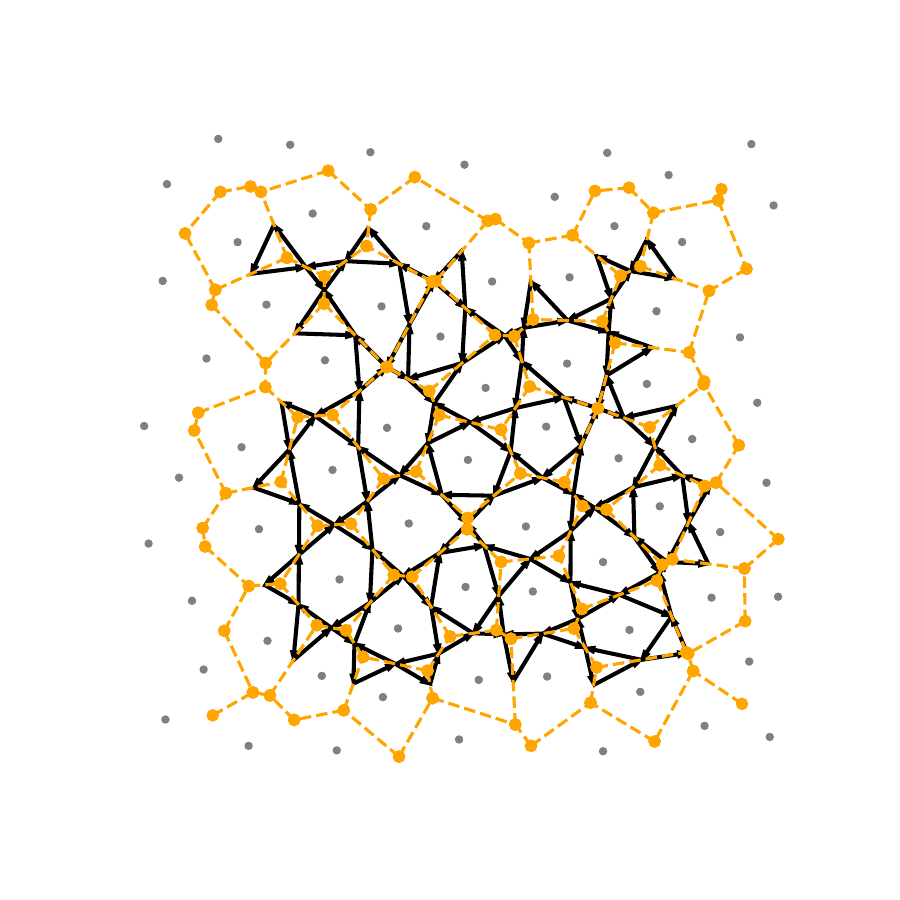}
\caption{Kagomization starts with a random, uncorrelated set of points (grey). Next, we generate the Voronoi diagram for this point set (orange) and then connect the midpoints on each edge to the neighboring edge midpoints to form the corresponding kagomized graph (black).}
\label{fig:kagomization}
\end{figure}

Beginning with an uncorrelated set of $L_0^2$ points in an $ L_0 \times L_0$ region we use the line graph construction \cite{mielke1991ferromagnetism} to generate a graph with the short range order defined above but no long range order.  Following \citep{mitchell2018}, we refer to this procedure as kagomization due to the local similarity between the resulting graphs and the kagom\'{e} lattice. This process can be consistently done for any system size $L_0$ which is necessary for extrapolating the results to the thermodynamic limit, see as in Sec. \ref{sec:ALoc}. The kagomization procedure is explained in Fig. \ref{fig:kagomization}: we first construct a Voronoi diagram from the uncorrelated set of points, and then generate triangles around each Voronoi vertex by connecting the midpoints of all Voronoi edges emanating from the vertex. These triangles form a kagomized graph with $S \sim 6 L_0^2$ sites. We define the system size as $L = \sqrt{S}$. To implement this graph with periodic boundary conditions, we perform the same procedure for a random point set on a two-dimensional torus. 

Kagomizations of Voroni graphs consist of corner-sharing triangles: this makes them a special type of hourglass graph. Some properties of these emerging hourglass graphs can be conveniently understood by considering the corresponding Delaunay triangulation, the dual graph of the Voronoi graph. Triangulation graphs are in one-to-one correspondence with hourglass graphs. At the same time, not all triangulations can be realized as Delaunay triangulation graphs \cite{DILLENCOURT1990283} and thus not all hourglass graphs will be generated by the procedure described in the previous paragraphs.

Any hourglass graph can be generated by a series of local updates to graph connectivity starting from the kagom\'{e} lattice. The local updates consist in restructuring a pair of adjacent triangles which correspond to diagonal flips \citep{diagonal-flips-bound} of the dual triangulation. The ability to construct an arbitrary graph in this way follows from the proven analogous result for triangulation graphs \citep{NEGAMI1994225, diagonal-flips-original}. Thus, considering all hourglass graphs gives the intriguing possibility to introduce amorphousness in a perturbative way. The simulation results in the next section pertain to hourglass graphs from kagomization, however, the exact results and arguments apply to any hourglass graph.

To write the model of Eq. \eqref{eq: ham} on a amorphous graph, we must be careful to only define hoppings around triangles that belonged to the original kagom\'{e} lattice. We refer to triangles of an amorphous graph that belonged to the original kagom\'{e} lattice as primitive triangles, and any cycles on the graph formed by connected primitive triangles as secondary polygons. On the kagom\'{e} lattice, all secondary polygons are hexagons, whereas on amorphous kagomized graphs, secondary polygons consist of $N$-sided polygons where $N = 3, 4, 5, 6, \cdots$. Writing our model on secondary triangles will give incorrect hopping terms. Thus, we define Eq. \ref{eq: ham} on an amorphous graph by summing over all \textit{primitive} triangles $\Delta_i$ of the graph. Numerically, this distinction is accomplished by writing the Hamiltonian from hoppings around the Voronoi vertices that give rise to a given kagomized graph. For example, to define the model on the graph shown in Fig. \ref{fig:kagomization}, we assign hoppings to all of the black edges. The signs the hopping phases are determined by their positions in the triangles of black edges that encircle the orange points.

\begin{section}{Compactly localized flat band states in amorphous model} \label{sec:flat}

The model on a kagom\'{e} lattice gives rise to three energy bands in momentum space, which are gapped or ungapped depending on the choice of $\phi$. A single flat band is present when $\phi =  \ell \pi/6$ for integer $\ell$. The degeneracy is at most $S/3 + 1$, where $S/3$ corresponds to the number of secondary hexagons in the graph.

The amorphous graph does not consist of only primitive triangles and hexagons. However, as in the kagom\'{e} lattice, every graph edge is a side of a triangle, and every vertex of the graph is tetravalent. Based on these observations, we show in App. A that flat band states always occur at energy $E_{fl} = 2\cos(3\phi + \pi)$. Below we describe in detail the structure of flat band states in amorphous systems for $\phi = 0$, and outline how this analysis can be generalized to $\phi = \ell \pi/6$. Additionally, we find that for other commensurate values of $\phi$ flat bands are only present in amorphous systems, not their kagom\'{e} lattice counterparts.  

\subsection{Robustness of flat bands to amorphousness in absence of field}

In the kagom\'{e} lattice at $\phi$ = 0, the local value (non-normalized) of a given flat band state $\ket{\psi_{fl}}$ alternates as $\pm1$ around a single hexagon and vanishes elsewhere, see Fig. \ref{fig:flat_band} (a). When the Hamiltonian acts on $\ket{\psi_{fl}}$, destructive interference on each site neighboring the hexagon prevents this state from dispersing outside of the hexagon. This flat band state has degeneracy $S/3 + 1$.

We conjecture and prove in App. A that a spanning set of all flat band states on an amorphous graph for $\phi = 0$ is given by two types of states. Type-I are defined by the following conditions:

\begin{enumerate}
    \item The state resides on a closed loop of even length.
    \item This loop incorporates an even number of vertices of each primitive triangle, which is equivalent to the loop being non-intersecting
    \item The state has (un-normalized) amplitudes $\pm 1$ which alternate along the loop.
\end{enumerate}

An example of a Type-I flat band state localized around a pentagon-septagon pair is shown in Fig. \ref{fig:flat_band}(b). Note that the wavefunction on the central site of this loop vanishes and that the action of the Hamiltonian on this state leads to destructive interference on this central site and all sites immediately neighboring the combined polygon. Type-I states can also reside on non-contractible loops of the torus. For periodic systems, the presence of non-contractible loops among flat band states was previously recognized in \cite{rhim2021singular}.
 
Type-II states consist of pairs of non-adjacent odd-length secondary polygons that satisfy the following conditions: 

\begin{enumerate}
    \item The state resides on a combination of two non-adjacent closed loops connected by a path
    \item The state has (un-normalized) amplitudes $\pm 1$ which alternate around the loops, but necessarily leaves two adjacent sites with the same sign, giving each loop a net sign
    \item The connecting path must start at the third vertex of the primitive triangle that contains two adjacent sites with the same sign on one polygon, and end at the analogous vertex of the other polygon. The state on this path has (un-normalized) alternating amplitudes $\pm 2$ 
\end{enumerate}

We refer to the connecting path as a ``Dirac string" in analogy to electromagnetism where a Dirac string connects two magnetic monopoles. The Dirac string is necessary for a Type-II state to be an eigenstate of Eqn. \ref{eq: ham} on an amorphous graph. In fact, the Dirac strings appear exclusively in flat band states on amorphous graphs of our model system! An example of a Type-II state is shown in Fig. \ref{fig:dirac_string_example}.

Numerical investigations find that the span of the set of Type-I and Type-II flat band states is \textit{exactly} one third of the number of sites $S$ in the graph. We also prove this result rigorously in App. A.

\subsection{Robustness of flat bands to amorphousness at non-zero field}

For $\phi = \frac{\ell}{6}\pi$, where $\ell$ is an integer, we similarly observe the persistence of a macroscopically degenerate levels (flat bands) in numerical simulations of amorphous systems. 

Analogous to the previous subsection, one can construct a macroscopic number of flat band states that generalize the flat band states localized on hexagons on the kagom\'{e} lattice. We see that some flat band states reside on combinations of multiple polygons in these amorphous graphs. To define these states, we employ a $\phi$-dependent integer $p$, whose $\ell$-dependence is explained in App. A, for example $\phi = \pi/6, \pi/3, \pi/2$ corresponds to $p = 3, 6, 1$.  To extend the conditions for Type-I flat band states to general $p$, we add $p$-dependence to conditions $1, 3$ : (1) flat band states must reside on loops whose length is divisible by $p$ ($p = 1$ implies any loop is valid) (3) the amplitudes are the $p$-th roots of unity, taken sequentially around the unit circle in the complex plane. For $\phi \neq 0$ it is also important to note that graphs become directed, and all loops that host flat band states must respect the orientation of the edges. 

Remarkably, for $p = 1, 2, 6$ which correspond to $\phi = \pi/2, 0, \pi/3$, we see that the fraction of all eigenstates that are in the flat band remains $1/3$ even on the amorphous graphs. In general we do not expect this to be the case. For example in Table \ref{table:flat_band_deg} we see that for $\phi =\pi/6$, the fraction of flat band states never exceeds $1/16$ in our simulations. Additionally, we observe that this fraction varies appreciably from realization to realization for most values of $\phi$, see  Table \ref{table:flat_band_deg}. The reasoning for why this does not occur for $p=2$ was explained in the previous subsection; the case $\phi = \pi/2$ ($p = 1$) can also be easily understood. The flat band state in this case does not alternate $\pm 1$ around a polygon but is in fact uniform. This simple structure of the flat band states ensures compatibility with all secondary polygons in the graph: every secondary $n$-gon in an amorphous system supports a uniform localized flat band state surrounding it, see Fig. \ref{fig:flat_band}d. We believe an analysis similar to the previous subsection can be performed in these cases except that one would need to consider directed graphs. We leave this for future work. 

\begin{figure}
    \centering
    \includegraphics[width=0.25\textwidth]{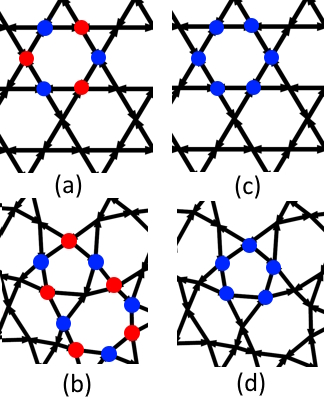}
    \caption{Figures on left side represent a flat band state at $\phi = 0$; figures on right side represent a flat band state at $\phi = \pi/2$. (a) Flat band eigenstate alternates between $+1$ (blue circles) and $-1$ (red circles) around the hexagon and is zero-valued outside. (b) Flat band eigenstate of an amorphous kagomized system wraps around a pentagon-septagon combination such that the state's sign can still alternate between circumferential sites. (c) At $\phi = \pi/2$ the flat band state has the same phase at every site surrounding a hexagon. (d) Since the state's sign is constant around a polygon, the pictured flat band state localized around a pentagon in an amorphous kagomized system is an eigenstate for $\phi = \pi/2$.}
    \label{fig:flat_band}
\end{figure}

\begin{figure}
    \centering
    \includegraphics[width=0.3\textwidth]{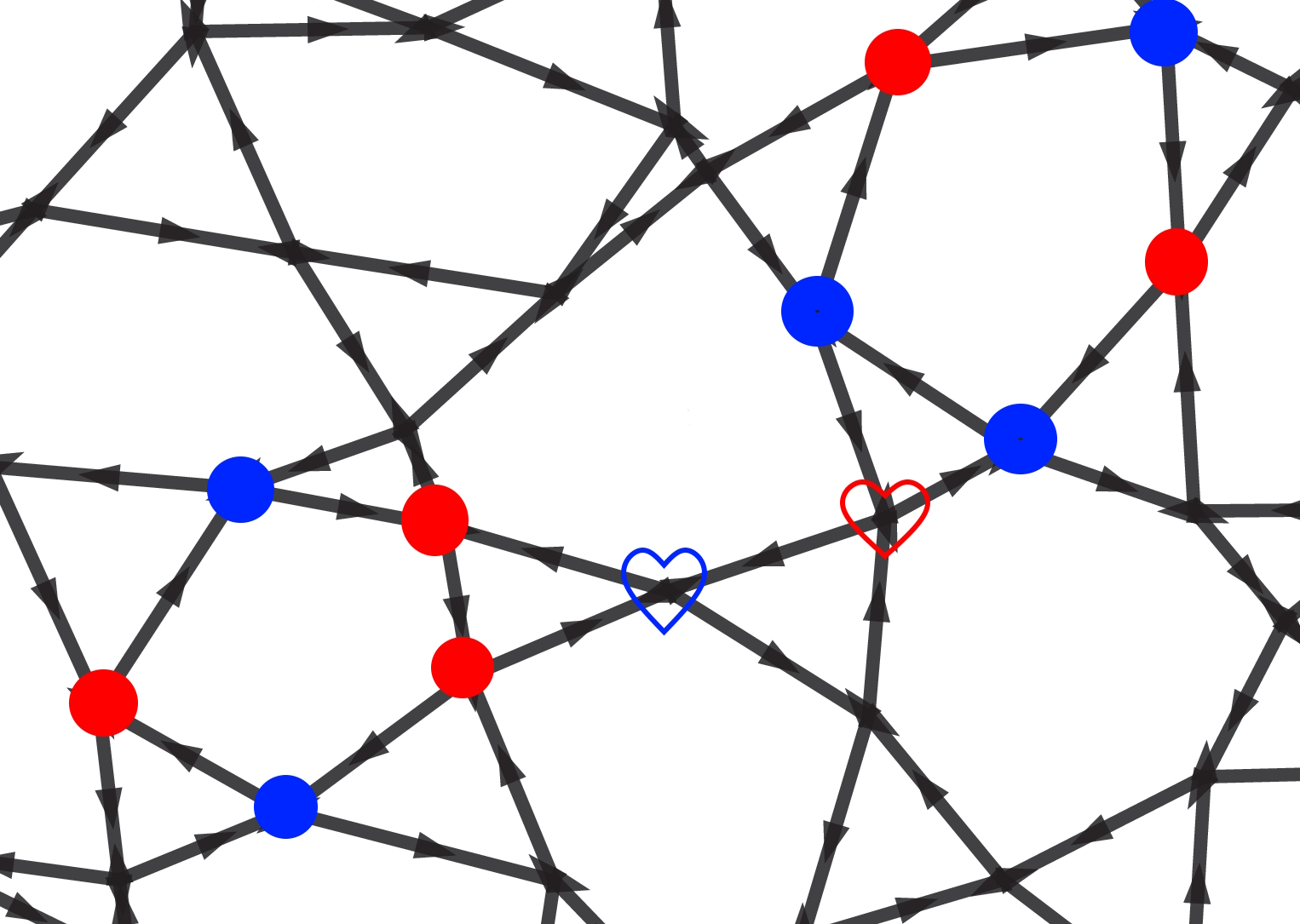}
    \caption{Example of a flat band state at $\phi = 0$ that combines two \textit{non-adjacent} polygons of odd length. The flat band eigenstate alternates between $+1$ (blue circles) and $-1$ (red circles), but leaves a net sign around each pentagon. While the compact localized state around each pentagon individually could not be an eigenstate, a ``Dirac string" of $+2$ (blue heart) and $-2$ (red heart) connects the pentagons resulting in a flat band eigenstate.}
    \label{fig:dirac_string_example}
\end{figure}

\begin{table}[]
    \centering
    \begin{tabular}{c|c|c}\toprule
         $\phi$ & kagom\'{e} lattice & amorphous system \\ \midrule
         0 & S/3 + 1 & S/3 \\
         $\pi/6$ & S/3 & $\leq$ S/16 \\
         $\pi/3$ & S/3 + 1 & S/3 \\
         $\pi/2$ & S/3 &  $\lesssim$ S/3 \\
         \hline 
         $3 \pi/8$ & 0 & $\leq$ S/16 \\
         $3 \pi/4$ & 0 & $\leq$ S/8 \\
         $11\pi/10$ & 0 & $\leq$ S/6
    \end{tabular}
    \label{table:flat_band_deg}
\caption{Numerical results (from 50 simulations of amorphous systems with $S = 48$) of flat band degeneracy for a subset of values of $\phi$ that give flat bands. Simulation results are shown for when the underlying graph of the system is a kagom\'{e} lattice or amorphous graph. Note that the number of flat band states can depend on the realization of amorphous disorder, in which case we provide an upper bound.}
\end{table}

\subsection{Flat bands unique to amorphous systems}
On the kagom\'{e} lattice, secondary hexagons cannot support localized eigenstates when $p$ does not divide $6$. However, amorphous graphs contain secondary polygons of various sizes, e.g. pentagons and septagons. Localized states around these polygons become possible through appropriately tuning $\phi$ to $\phi = \ell\pi/p  - \pi/2$ for integer $\ell$. This result can be derived from the results of App. A. We conjecture that at these values of $\phi$, flat bands that were not present in the kagom\'{e} lattice will appear in amorphous systems! In Fig. \ref{fig:flat_band_amorphous_only}, we illustrate it for the case $\phi = 11\pi/10$, corresponding to $p = 5$ (states localized around pentagons).
\begin{figure}
    \centering
    \includegraphics[width=0.48\textwidth]{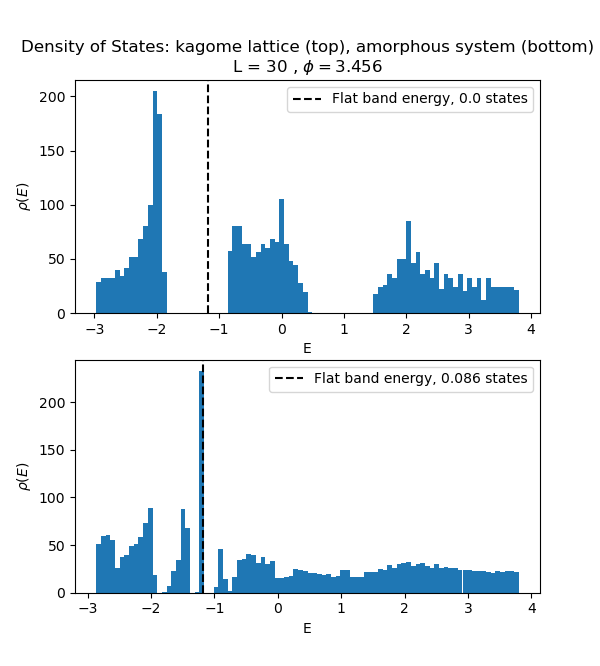}
    \caption{Example of a case where a flat band exists in the amorphous system but \textit{not} in the kagom\'{e} lattice system. For $\phi = 11\pi/10$, a flat band emerges in the amorphous system in the gap of the kagom\'{e} lattice. This flat band contains $.086$ of the total states of the system, unlike the $1/3$ typically observed in flat bands in these systems. We hypothesize that these flat band states are hosted on pentagons.}
    \label{fig:flat_band_amorphous_only}
\end{figure}

We note that these flat bands are expected to be topologically trivial. In $2D$, it has been proven that topologically nontrivial flat bands can only exist if hopping is non-local\citep{chen-Chern-nope}, which is not the case in our model. Though the proof was done for periodic systems, we expect a similar result would apply in our case due to the local nature of most states in the flat band.

\subsection{Other lattices}

Besides the kagom\'{e} lattice, there are multiple other lattices that can support flat bands \cite{neves2024crystal}. We expect that the fate of flat bands in the amorphous versions of these lattices will be similar to what we described for the kagom\'{e} case. As an illustration in Fig. \ref{fig:lieb}, we show how localized states can persist in the amorphous version of the Lieb lattice. Other lattices that we believe will host flat band states in the amorphous form include the  star lattice and dice lattice. 

\begin{figure}
    \centering
    \includegraphics[width=0.7\columnwidth]{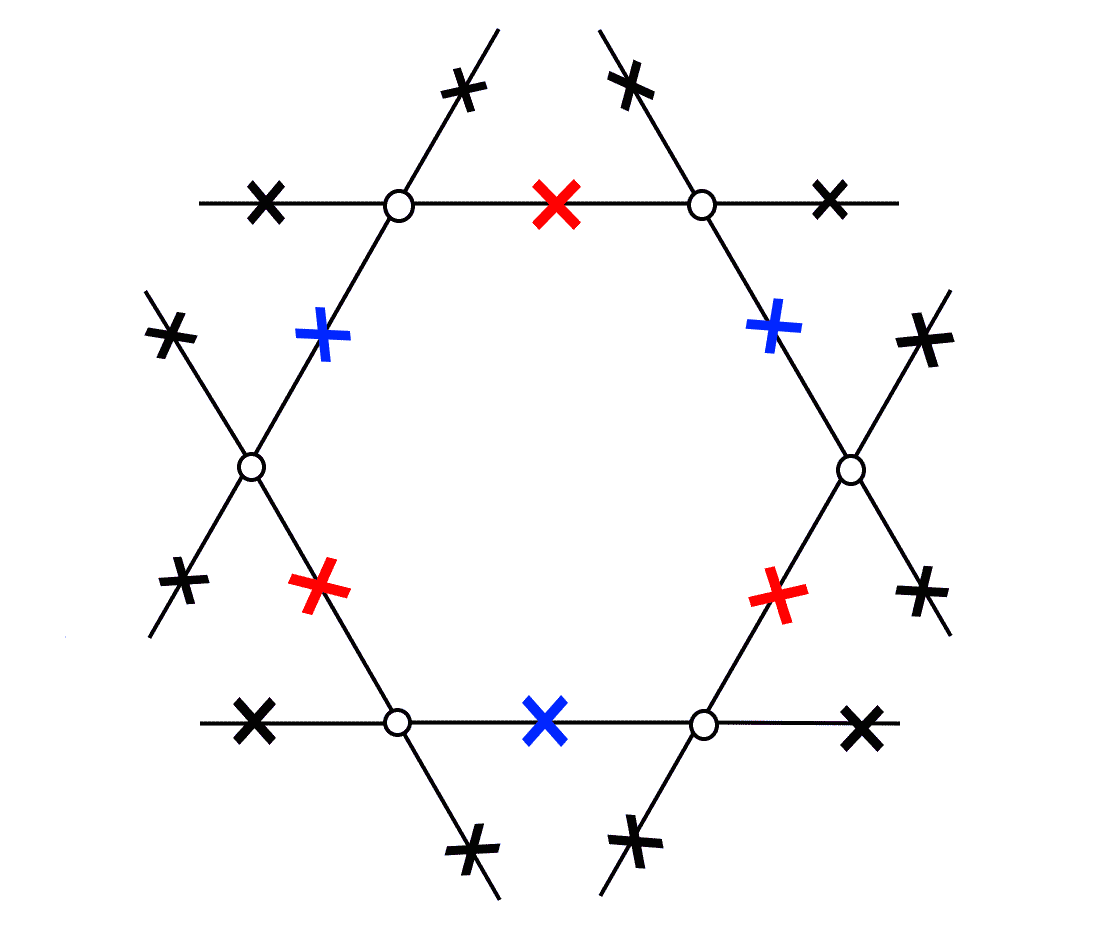}
    \caption{Example of a localized state on an amorphous analog of the Lieb lattice. In the Lieb lattice, states can exist both at vertices of the graph as well as at midpoints of the graph edges, denoted by $X$ in the figure. Blue (red) $X$ denote non-normalized state amplitude $+1$($-1$) whereas white circles at the vertices and black $X$s denote no state amplitude.}
    \label{fig:lieb}
\end{figure}

We will now focus on the dispersive bands of our model which, in the amorphous case, exhibit a distinct type of localized states that we refer to as Anderson-like localized states.
\end{section}

\begin{section}{Anderson-like localization in amorphous systems} \label{sec:ALoc}

In a previous work including the authors of this study \citep{zhetao-amorphousnl}, we observed localization in the dispersive bands of amorphous systems in class D. The localization of states was dependent on their eigenenergy, with energy regions in the band tails giving rise to localized states. In this work, we confirm this observation by adapting two methods previously used to study localization: effective localization-length spectra and nearest energy level spacings. We then compare our results in class D (Sec. \ref{sec:class_D}) to class A, accessed by tuning $\phi$ (Sec. \ref{sec:classA}).

We refer to the localization we observe as ``Anderson-like" to distinguish it from true Anderson localization. Anderson localization is defined by states whose magnitude decays from a well-defined center as $|\psi|^2 \sim e^{-r/\xi}$ where $r$ is the distance from the center and $\xi$ is the localization length \citep{Mott-Allgaier}. In Anderson-like localized states, the state can be localized about multiple centers, but still decays rapidly away from each center. Anderson-like localized states were previously observed in amorphous systems hosted on Voronoi graphs \citep{thesis-martin}. To detect localization of a state $\psi$, we calculate its inverse participation ratio $I_2$ (IPR), defined as
\begin{equation}
    I_2 = \sum_{i = 1}^{S} |\psi_i|^4
    \label{eq:def_ipr}
\end{equation}

Where $S$ is the number of sites on the graph, $\psi_i$ is the amplitude of the state on site $i$, and $\psi$ is normalized. We define an effective localization length $\tilde{\xi} = \frac{1}{\sqrt{I_2}}$. While inequivalent to the true localization length $\xi$, best approximated using the smallest quasi-1d Lyapunov exponent \citep{mackinnon-kramer}, the effective localization length is well suited to describing localization in amorphous systems for systems sufficiently small to be efficiently diagonalized. 

We also analyze nearest-level spacings distributions, \ejd{adapted for these systems. In the main text we give the results, specifically whether the distribution fits a Poisson law (localized states), a Wigner surmise (delocalized states), or neither.  Details of these analyses are described in App. \ref{app:lsd}.} 

In the $2D$ Anderson model all eigenstates of a system are localized for arbitrarily small onsite disorder strength \cite{girvin2019modern}. For large onsite disorder the $3D$ Anderson model also becomes insulating at all energies, however for sufficiently small onsite disorder, energy eigenstates at low energies are delocalized and energy eigenstates at higher energies are localized, separated by a mobility edge \cite{3d-Anderson}. We observe that amorphous systems qualitatively resemble the $3D$ Anderson model with weak disorder: the states can be either localized and/or delocalized, depending on which energy we probe. 

\begin{subsection}{Anderson-like localization in class D}\label{sec:class_D}

We start with a brief overview of the spectral properties of the periodic (kagom\'{e} lattice) system in class D ($\phi = \pi/2$). This system has three gapped energy bands in momentum space: a flat band at $E = 0$ sandwiched by two particle-hole symmetric bands $E_{\pm}$. We focus on the highest energy band throughout all following analyses. The top band spans $E = (\sqrt{3}, 2\sqrt{3})$ and has a Lifshitz transition at $E = 2$. Due to momentum space degeneracy, $I_2(E_{+})$ is multi-valued and $\tilde{\xi}(E)$ forms an envelope. This envelope is strongly centered around $\tilde{\xi}(E) \sim O(1)$ for all energies, consistent with the delocalized nature of these eigenstates that we expect because of translation invariance.

Amorphous class D systems retain the gapped three-band structure of their kagom\'{e} counterparts. The flat band is protected from amorphousness by particle-hole symmetry as well as the arguments outlined in Sec. \ref{sec:flat}. Numerically, we generate and diagonalize the Hamiltonian of amorphous systems of size $L_0$ ranging from $L_0 = 40$ to $L_0 = 100$ and calculate $I_2$ for all eigenstates.  To focus on bulk properties, we apply periodic boundary conditions in both directions for all calculations, unlike in \citep{zhetao-amorphousnl} where we used open boundary conditions.

The effective localization spectrum for a large class D amorphous system ($L_0$ = 100) is shown in Fig. \ref{fig:D_spectrum}. The density of states of the bulk band extends into the bandgap of the kagom\'{e} lattice but the amorphous disorder is not sufficiently strong to close the gap. Most states in the amorphousness-induced tail, with energies $E < \sqrt{3}$, have suppressed effective localization length. However, not all states with $\tilde{\xi} << L$ are in the tails: even in large systems, some states deep in the metallic (delocalized) region of the bulk are relatively localized. In fact, effective localization length $\tilde{\xi}$ can vary considerably between states that are close in energy, forming an envelope structure similar to that observed in the kagom\'{e} lattice. The envelope structure complicates defining a mobility edge in these systems. To approximate a mobility edge, we define energy ranges such that the effective localization length probability distribution can be split into two distinct regions, the ``localized" states, shown in blue in Fig. \ref{fig:D_spectrum} and ``delocalized" states shown in red, with minimal overlap. All eigenstates outside of these categories we categorize as \ejd{``intermediate"} states (purple). Setting cutoff $\tilde{\xi}/L < 1/3$ for classifying localized states coincides with this analysis. 

\begin{figure}
    \centering \includegraphics[width = .7\columnwidth]{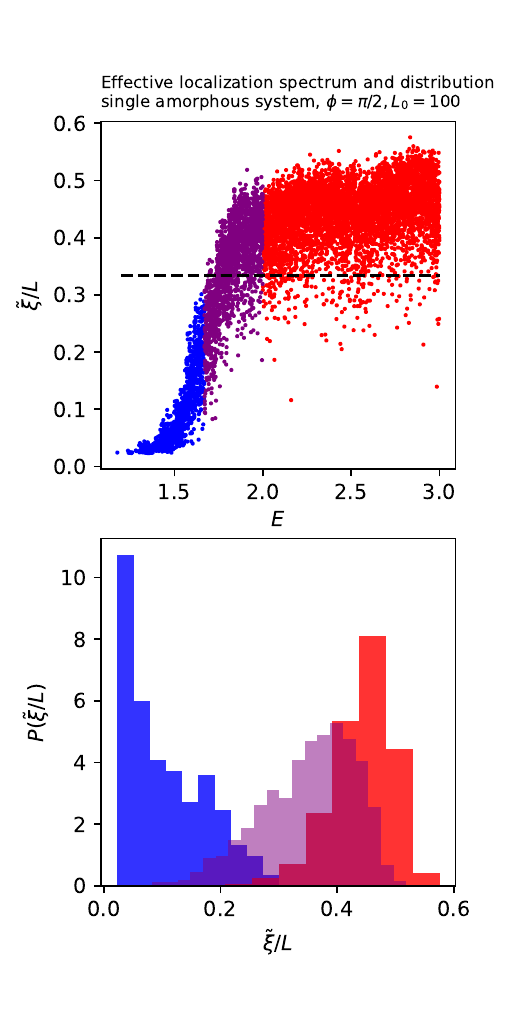}
    \caption{Top: Scatter plot of the energy of every eigenstate in the highest energy band of an $L_0 = 100$ amorphous system and its effective localization length. States are separated into three energy ranges, corresponding to localized (blue), \ejd{intermediate} (purple) and delocalized (red). Bottom: States classified as localized (red) have nonzero probability $P(\tilde{\xi}/L)$ for $\tilde{\xi}/L < 1/3$. The energy range for delocalized states is set such that the delocalized states' (red) probabilities' non-zero range has minimal overlap with the localized states' range. All states in-between are classified as \ejd{intermediate} states, whose distribution $P(\tilde{\xi}/L)$ is shown in purple.}
    \label{fig:D_spectrum}
\end{figure}

Nearest-level-spacings analysis of states assigned to the localized energy range fit a Poisson law well. The spacings of states assigned to the delocalized energy range do not come from a Poisson law distribution, nor do they fit the Wigner distribution with a high degree of certainty. \ejd{These distributions are shown in Fig. \ref{fig:D_deloc_ess} in App. \ref{app:lsd}}. This inconclusivity is likely due to the persistence of localized states deep in the metallic phase, as shown in Fig. \ref{fig:D_spectrum}.

Considering a single realization of amorphousness at experimentally relevant sizes without scaling illustrates the complexity of localization in these amorphous systems. Scaling-based evidence for localization from amorphousness was previously shown in \citep{zhetao-amorphousnl} where we averaged $I_2$ over multiple realizations of amorphousness for various system sizes. We observed that $I_2$ for eigenstates in the bulk band tails remains approximately constant while increasing system size. While numerical data for only a single system size is presented here, additional data suggests that the approximate mobility edge location is independent of system size for systems from $L = 20$ to $L = 100$.

\end{subsection}

\begin{subsection}{Anderson-like localization in Class A}
\label{sec:classA}

In the following, we repeat the analysis of Sec. \ref{sec:class_D} for a system of Eq. \eqref{eq: ham} for a system with broken particle hole symmetry: in AZ class A. For subsequent calculations we set $\phi = 4\pi/9$, such that the bulk remains gapped, see Fig. \ref{fig:A_spectrum}. The middle band has now broadened and the surrounding bands are no longer particle-hole symmetric. The upper band, however, retains qualitatively similar features to the class D case. In Fig. \ref{fig:A_loc_ess} in App. \ref{app:lsd} we repeat the level spacings analysis of the previous section of subsets of this upper band and find qualitatively similar results to the analysis in class D. This agreement of localization properties between models in class A and class D with $E \neq 0$ was previously found and reasoned in a study of the thermal metal phase \cite{cosma-thermal-metal}, however in those models, that did not involve amorphous disorder, all states were localized.

\begin{figure}
    \centering
    \includegraphics[width = .8\columnwidth]{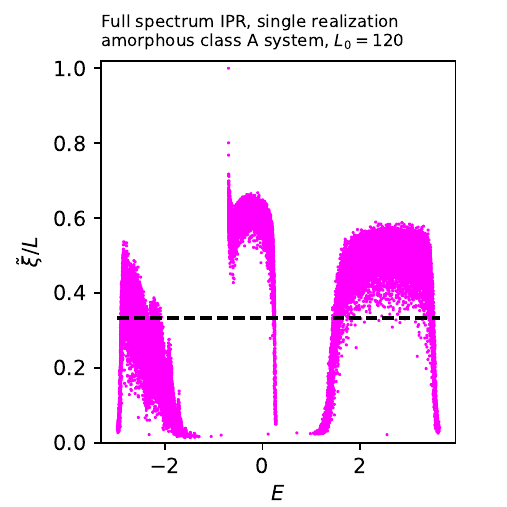}
    \caption{Scatter plot of the energy of every eigenstate of $L_0 = 120$ amorphous system with $\phi = 4\pi/9$, symmetry class A. The highest energy band localization is similar to highest energy band (and its particle-hole-symmetric partner) in class D. However, the flat band has broadened and the lowest band appears to have mostly localized states, unlike in class D. A fully delocalized state with $\tilde{\xi}/L = 1$ exists at $E = -4\cos(4\pi/9)$.}
    \label{fig:A_spectrum}
\end{figure}

For context, we review current thinking about which properties are universal in generically disordered systems, using class A systems as an example. This model can be mapped to a $2D$ disordered Dirac fermion picture. A long-standing conjecture suggested that the localization-delocalization transition of $2D$ disordered Dirac fermions is controlled by the same fixed point as the integer quantum Hall transition of ordinary fermions\citep{ludwig1994}. In the integer quantum Hall transition, a class A system, the localization length of states in a bulk band diverges at the critical energy $E_c$ as $\xi \sim |E - E_c|^{-\nu}$ \citep{huckestein1995}. Thus in class A disordered Dirac fermions, we expect states in the bulk bands to be fully localized even for infinitesimally weak disorder except for a single delocalized state in the center of the bulk band. Numerical results have confirmed this fact for onsite disorder in class A \citep{sbierski2020c}, even if it is not absolutely clear that the exponent $\nu$ is the same between Dirac and quadratic fermions in a magnetic field. We cannot interpret our results for amorphous disorder in class A within this basic picture due to the extended range of delocalized states shown in Fig. \ref{fig:A_deloc_ess}. This suggests that the amorphous nature of the system may require an expansion of the symmetry classification to include new types of localization phenomena.

We stress that we cannot strictly rule out the possibility that an extended range of delocalized states for amorphous class A systems is a finite size effect. To evaluate this claim, one must numerically compute the conductance $g$ and mean free path $\ell$ for amorphous systems, compare $\ell$ to the system sizes we study, and $g$ to the criteria derived in \cite{ostrovsky-2015}. We note, however, that the system sizes we simulate are on the scale of experiments with meta-materials, thus even if delocalized states only appear so as a result of finite size, they may be effectively metallic for applications. 

\end{subsection}

\end{section}

\begin{subsection}{\ejd{Localization-delocalization transition unique to amorphous systems}}
\label{sec:ldl}

\ejd{In the class A amorphous system (with $\phi = 4\pi/9)$, we observe a localization feature not seen in systems with onsite-disorder: a completely delocalized state, here at the left edge of the middle band ($E = -.695$), see Fig. \ref{fig:A_spectrum}. This uniform state, with equal amplitude on each site, is an eigenstate of Eq. \eqref{eq: ham} with energy $E = -4\cos{\phi}$. The fully delocalized state did not appear in the class D analysis because it occurs at the same energy as the flat band states ($E = 0$) and can be written as a superposition of flat band states. The class A amorphous system we consider has no flat band. 

The existence of this completely delocalized eigenstate is a consequence of the fixed coordination number in these graphs. For the uniform state, the local Hamiltonian appears the same on every site of the system and this state will occur in systems of all sizes. Remarkably, we observe that effective localization lengths of states nearby in energy to the fully delocalized state are significantly augmented, suggesting the existence of a localization-delocalization transition with no analog in generic disordered systems.

Other examples of delocalization at an isolated energy with a diverging localization length of localized states nearby are the 1D chain with random nearest-neighbor hopping and the 2D quantum Hall plateau transition, but in the present model the delocalized state arises as a result of fixed coordination number rather than symmetry (as in those cases). We generally expect that in models with a short-range Hamiltonian (such as Eq. \ref{eq: ham}) and fixed coordination number, a fully delocalized state and associated localization-delocalization transition exist if the energy of the delocalized state does not coincide with that of a flat band.

\begin{figure}[htp]
\begin{subfigure}{}
\includegraphics[clip,width=.8\columnwidth]{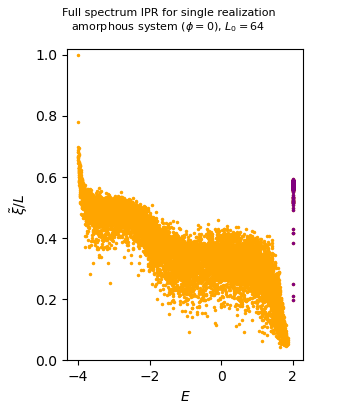}%
    \label{fig:B0_spectrum}
\end{subfigure}
\begin{subfigure}{}
\includegraphics[clip,width=0.8\columnwidth]{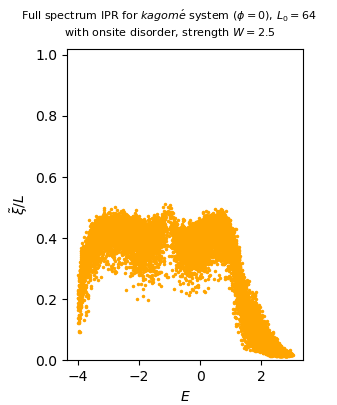}
  \label{fig:onsite}
\end{subfigure}
\caption{\ejd{Top: Scatter plot of the energy of every eigenstate of an $L_0 = 64$ amorphous system with $\phi = 0$ and its effective localization length. At $E = -4$ is a completely delocalized state and at $E = 2$ is the system flat band, colored in purple. Note that the localization length increases rapidly to the system size as $E \rightarrow -4$, and that the localization length is not uniquely defined for the massive degeneracy from flat band states at $E=2$. Bottom: Scatter plot of the energy of every eigenstate of an $L_0 = 64$ kagom\'{e} system with $\phi = 0$ and onsite disorder of strength $W = 2.5$, and its effective localization length.}}
\end{figure}

As another example, in Fig. \ref{fig:B0_spectrum}, we see this delocalization-localization in amorphous systems with $\phi = 0$, which belong to symmetry class AI. Near the fully delocalized state at $E = -4$, states become increasingly delocalized in character.  We compare with systems with onsite-disorder but no amorphous disorder in Fig. \ref{fig:onsite}. We set the disorder strength at $W = 2.5$, adding a randomized energy to each site between $[-W/2, W/2]$. In the onsite disorder case, equivalent to the 2D Anderson model on a kagom\'{e} lattice, we see no fully delocalized state or localization transition. 

The nature and critical properties of this localization-delocalization transition are interesting venues for future work. These observations, as well as the possibility of an extended range of metallic states in class A, also call for further study, both numerical and analytical, to expand the AZ symmetry classification to include amorphous disorder.}

\end{subsection}

\begin{section}{Conclusion}

In this work, we explored two types of localization in an amorphous topological-photonics-inspired model: compact localized states in flat bands and Anderson-like localized states. This model can access different physics and symmetry classes through tuning effective magnetic flux $\phi$. 

We show that flat bands appearing in the kagom\'{e} lattice model for $\phi = n\pi/6$, which are destroyed by onsite disorder, persist in the presence of amorphous disorder. This work opens up the study of features such as band-touching between flat bands and dispersive bands \citep{balents-band-touching} in amorphous systems. We also show that partial flat bands occur in amorphous systems at some values of $\phi$ which are \textit{absent} in their kagom\'{e} lattice counterparts. In these flat bands unique to amorphous systems, it is possible to tune the number of flat band states by modifying the structure of the underlying graph, which may be useful for some applications.

Beyond these localized flat band states, we also demonstrate Anderson-like localization that is induced by amorphousness. To study these localization properties, we use and adapt two methods: inverse-participation-ratio (IPR) calculations and energy-level-spacings analysis. We find that this type of localization occurs in the symmetry classes we would expect from studies of onsite disordered systems, however, the observed extent of the spectrum which is localized, particularly in symmetry class A, differs from the Anderson paradigm of localization.

For context for our results, we point out the work by Puschmann who considered random $2D$ Voronoi Delaunay lattices with out-of-plane uniform magnetic field (class A) and found quantum Hall edge states with localized bulk modes. This work also presents evidence that the critical properties of this model agree with models in class A with only onsite disorder \citep{thesis-martin}. Although this model is in class A, the implementation of amorphous disorder and magnetic field differed. This might mean the implementation of amorphousness is crucial for Anderson localization.

For additional context, one can also consider the Aubry-Andr\'{e} (AA) model, a 1D lattice with quasi-periodically modulated hoppings. Although universality predicts that all state must localize for an arbitrary weak disorder in the relevant symmetry class, in the AA model localization happens at a finite value of the modulation amplitude. This result suggests that in order to modify universality, disorder must be constrained to an extreme degree (there are only two global parameters in the case of the AA model, so arguably ``disorder'' is no longer even the right word). Here is an important distinction between our observation of the modification of universality and the analogous case in the AA model: our example exhibits refined universality in a model where the randomness still has an extensive number of degrees of freedom.

The methods we use to study localization in amorphous systems may also be applicable to studying newly-discovered fractional states in amorphous Chern insulators \citep{kim-amorphous-fractionalization} and amorphous chiral spin liquids \citep{cassella-amorphous-qsl}. There is a considerable body of work on how flat bands in electronic systems modify interacting electron physics~\cite{EPJB2001,PhysRevLett.88.167207,IJMPB2015}, which has expanded considerably with the advent of moire flat bands; while control of photon-photon interactions was one reason for experimental studies of amorphous photonic systems, further investigation of interactions in the photonic case is also needed. We also emphasize that our model system offers an approach to study introducing amorphousness perturbatively through combining local diagonal flips in the dual graph of a system. 

This work also has application to the study of phases and universal behavior in two-dimensional disordered topological superconductors in symmetry class D. Such superconductors show metal-insulator transitions with onsite disorder \citep{wang2021} and non-linear sigma model RG analysis predicts quasiparticle localization \citep{senthil2000}. While a recent study claims that amorphous Chern insulators in class D have non-universal critical behavior \citep{ojanen-classD}, more research is needed to verify these claims. Here we have proposed an alternate microscopic model, with the ability to tune the degree of amorphousness without destroying local order. This will be a useful tool to further study metal-insulator transitions in both topological superconductors and in amorphous Chern insulators in class D.

This work raises a multitude of questions that need further exploration, about (1) flat bands, (2) the fate of periodic band structures when the underlying graph is made amorphous and (3) Anderson-like localization in the thermodynamic limit. We outline these areas for continued investigation below.

\begin{itemize}
    \item What determines the degeneracy of a flat band? In App. A we derive the degeneracy for $\phi = 0, \pi/2$ for both kagom\'{e} and amorphous systems. However, as seen in Table \ref{table:flat_band_deg}, numerical simulations show that other values of $\phi$ have different flat band degeneracies for which our derivation is not sufficient.
    \item Why do some bands of periodic band structures survive while others do not when the underlying structure is made amorphous? For example, in our model systems we see that the class D three-band structure from the periodic case is preserved (with tails extending into the periodic system bandgap as in Fig. \ref{fig:D_spectrum}). However, for $\phi = 11\pi/10$ the density of states plots in Fig. \ref{fig:flat_band_amorphous_only} indicate that the lowest energy band retains its qualitative character, two additional bands (one flat) are created in the bandgap, and the highest two energy bands blur together. Further work is needed to explain why (a) some bands are minimally affected by the system becoming amorphous and (b) this effect depends on $\phi$.
    \item \ejd{How do we understand the envelope structure observed in the effective localization spectrum in amorphous systems? How do we reconcile the coexistence of localized and delocalized states at arbitrarily close energies in Fig. \ref{fig:D_spectrum} with Mott's argument, which forbids such coexistence in generic disordered systems? }
    \item How do we identify universality classes of amorphous localization phase transitions? What aspects of amorphous disorder become relevant to such a classification?  
\end{itemize}

\end{section}

\begin{section}{Acknowledgements}
    We gratefully acknowledge helpful discussions with Martin Puschmann, Thomas Vojta, Mert Bozkurt, Anton Akhmerov, Ilya Gruzberg, Adolfo Grushin, Omer Mert Aksoy and Alexander Tsigler. E.J.D was supported by the NSF Graduate Research Fellowship Program, NSF Grant No. DGE 1752814. A.A. and J.E.M. were supported by the NSF QLCI program through Grant No. OMA-2016245. Additionally, A.A. was supported by a Kavli ENSI fellowship during his time at UC Berkeley and National Science Foundation (NSF) Convergence Accelerator Award No. 2235945 at MIT. Computations were performed at the Lawrencium cluster at Lawrence Berkeley National Lab.
\end{section}

\appendix

\section{Flat Band calculations}
\label{app:flat}

\subsection{Flat band structure}
\begin{figure}[t]
\centering
\includegraphics[width = 0.2\textwidth]{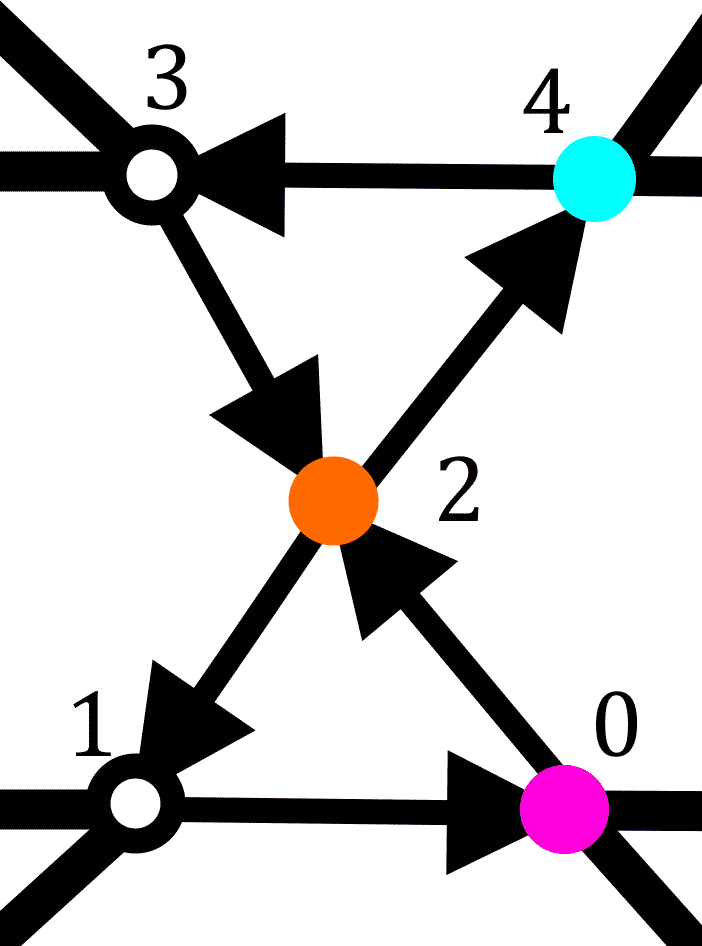}
\caption{A flat band state on a segment of an $n$-gon. The flat band state is valued $e^{i\alpha}$ (pink circle), $e^{i\beta}$ (orange circle), and  $e^{i\gamma}$ (teal circle) on sites inside the polygon, and vanishes outside the polygon (white circles). The Hamiltonian, particularly the choice of $\phi$, determines the relative phases of the state around the polygon, and the graph geometry determines whether the system hosts a flat band eigenstate at this value of $\phi$.}
\label{fig:flb}
\end{figure}

We can construct these flat band states for a Hamiltonian of general phase by considering the action of the Hamiltonian on a flat band state defined around a single section of an $n-$gon. In Fig. \ref{fig:flb} we define a basis such that even-indexed sites form a polygon and odd sites are connected to the polygon. Note that this construction applies to any polygon of $n \geq 3$. Based on this diagram, we formulate an ansatz for a (non-normalized) flat band state as $|\psi_{fl}\rangle = e^{i\alpha} |0\rangle + e^{i\beta} |2\rangle  + e^{i\gamma} |4\rangle + \cdots$ where the state vanishes on odd-index sites in the basis and exists on even-indexed sites of the $n-$gon peripheral to this diagram. Action of the Hamiltonian gives:  

\begin{align*}
    H |\psi_{fl}\rangle &= (e^{i\alpha}e^{-i\phi} + e^{i\beta}e^{i\phi}) |1\rangle + \\
    & (e^{i\alpha}e^{i\phi} + e^{i\gamma}e^{-i\phi}) |2\rangle  + (e^{i\gamma}e^{i\phi} + e^{i\beta}e^{-i\phi}) |3\rangle + \cdots
\end{align*}

For $|\psi_{fl}\rangle$ to be an eigenstate:

\begin{align}
\Delta \phi = \alpha - \beta = \beta - \gamma \mod{2\pi} &= 2\phi + \pi \mod{2\pi} \\
e^{i\alpha}e^{i\phi} + e^{i\gamma}e^{-i\phi} &= E_{fl} e^{i\beta}
\label{eq:flb}
\end{align}

Combining the above equations gives:

\begin{equation}
    E_{fl} = 2 \cos{(3\phi + \pi)}
\end{equation}

This analysis suggests that, in principle, a flat band state could exist at any value of $\phi$ in our model systems. However, the geometry of the underlying graphs of these systems restricts the values of $\phi$ at which we see flat bands. States can only be eigenstates of the Hamiltonian if they live on $n$-gons where $ n\Delta \phi = 0 \mod{2\pi} $ and $\phi$ determines the value of $\Delta \phi := \alpha - \beta = 2\phi + \pi \mod{2\pi}$. The integer $p$ used in the main text to find loops on amorphous graphs that can host flat bands is found by counting how many iterations of $\Delta \phi$ are required before the phase returns to a multiple of $2\pi$.

Numerical diagonalization of kagom\'{e} systems reveals that flat bands only exist when $\phi$ is a multiple of $\pi/6$. For the kagom\'{e} lattice, hexagons are the only secondary polygons and thus $6(2\phi + \pi) = 2\pi m$ for $m \in \mathbb{Z}$ . This condition indeed gives that $\phi$ must be a multiple of $\pi/6$ to observe a flat band. Note that flat band eigenstates never occur localized around the primitive triangles of the kagom\'{e} lattice because this geometry does not lead to destructive interference when the Hamiltonian acts on such a state.

An amorphous kagomized graph, however, contains secondary polygons with $n\neq 6$ in addition to hexagons. The following condition determines the possible phases $\phi$ for observing flat bands in amorphous systems:

\begin{equation}
n\Delta \phi = n(2\phi + \pi) = 2\pi m \hspace{.25cm} \forall m \in \mathbb{Z}
\label{eq: great_plains}
\end{equation}

This condition reproduces the flat bands seen in kagom\'{e} systems and predicts values of $\phi$ that only show flat bands in the amorphous case. An example of an amorphous-only flat band is seen at $\phi = 11\pi/10$ for $n = 5$, $m = 8$, the flat band shown in Fig. 4 in the main text. 

\subsection{Flat band degeneracy}
In the previous section, we saw that flat band states are intimately connected to cycles on directed, kagomized graphs. The precise relation depends on the value of the flux $\phi$. Here we will work out this relation for the case of $\phi = 0$ where one does not need to specify the direction on the graph (since the magnetic field is absent).

First, let us establish some basic properties of kagomized graphs. We consider a graph with $S$ sites. The number of primitive triangles $N_{\Delta} = \frac{2}{3} S$, and, as a consequence, $N_{\Delta}$ is always divisible by 2 and $S$ is always divisible by 3. The number of edges is $3 N_{\Delta}$. Next, we would like to point out, that secondary polygons are well defined only when we consider graph's embedding. In what follows, we will assume that the graph is embedded on a torus, as has been the case for the numerical simulations in this paper. Euler's formula for the torus topology reads: 

\begin{equation}
S - 3 N_{\Delta} + (N_{\Delta} + N_{poly}) = 0,
\label{eq:euler}
\end{equation}

where $N_{poly}$ is the number of secondary polygons in the graph. We conclude $N_{poly} = S/3$ (it would be $N_{poly} = S/3 - 1$ in the case of disk topology).

Lastly, if there are odd polygons in the graph, their number is always even. This following from the fact that the total number of edges in all polygons is given by $3 N_{\Delta}$, and we have already established that $N_{\Delta}$ must be even.

From Table 1, we see that the flat band degeneracy never exceeds $S/3+1$. This can be interpreted as the number of independent cycles that contain an even number of vertices (i.e. 0 or 2) of each primitive triangle, we denote it as $C$. For the most part these cycles just come from secondary polygons, but we need to account for topology. We note that on a torus, the sum of cycles on all polygons vanishes, additionally, there are two non-contractible cycles. Overall, with Eq. \ref{eq:euler}, this yields

\begin{equation} \label{eq:even_count}
C = S/3 -1 + 2 = S/3 + 1.
\end{equation}

The answer would be the same if we assumed disk topology. We expect the flat band to derive from these cycles in some way. In the rest of the section, we will perform this construction for the case $\phi=0$. 

\textbf{The case of $\phi = 0$:} Recall, that the Hamiltonian Eq. 1 is given by a sum of terms supported on individual primitive triangles, each of which we will refer to as a local Hamiltonian. Each term has form $|1\rangle \langle2| + |2\rangle \langle3| + |3\rangle \langle1| + h.c. $ with eigenvalues $2,-1,-1$. Each flat band state has energy $-2$ and this is only achievable if the state lies in the $-1$ subspace of the local Hamiltonian for each primitive triangle, since each site is acted upon by exactly two terms.

The $-1$ subspace is spanned by states that have amplitudes $\pm 1$ on the sites of any edge in the primitive triangle. Requiring that the amplitudes sum to zero on every primitive triangle leads us to conclude that the space of flat bands is equivalent to the space spanned by states supported on loops with the amplitude alternating  $\pm 1$ along the loop. We note that a loop is permitted to go over the same edge multiple times, as occurs on the edges of Dirac strings in Type-II states introduced in Sec. III. Thus, a Type-II state is also considered a loop in this construction. 

Finally, we can compute the number of such states on a kagomized graph with $S$ sites. If all secondary polygons are of even length, then the total flat band subspace spanned by the alternating states on each polygon plus the alternating states on non-contractible loops (their length can always be made even by making them use two edges of a triangle instead of one or visa versa). This is the case for periodic kagom\'{e} lattice and the number states coincides with Eq. \eqref{eq:even_count}: $S/3+1$. In the case when there are odd secondary polygons (they must come in pairs), we built a flat band state by combining any pair polygons pairwise if adjacent (Type-I states in Sec. III) or connecting them with a Dirac string otherwise (Type-II states in Sec. III). If there are $M$ odd secondary polygons, this procedure will yield $M - 1$ linearly independent flat band states, leading to the degeneracy of $S/3$ confirmed by numerical simulations in Table 1.

\begin{figure}[htp]
\begin{subfigure}{}
    \includegraphics[clip,width=.9\columnwidth]{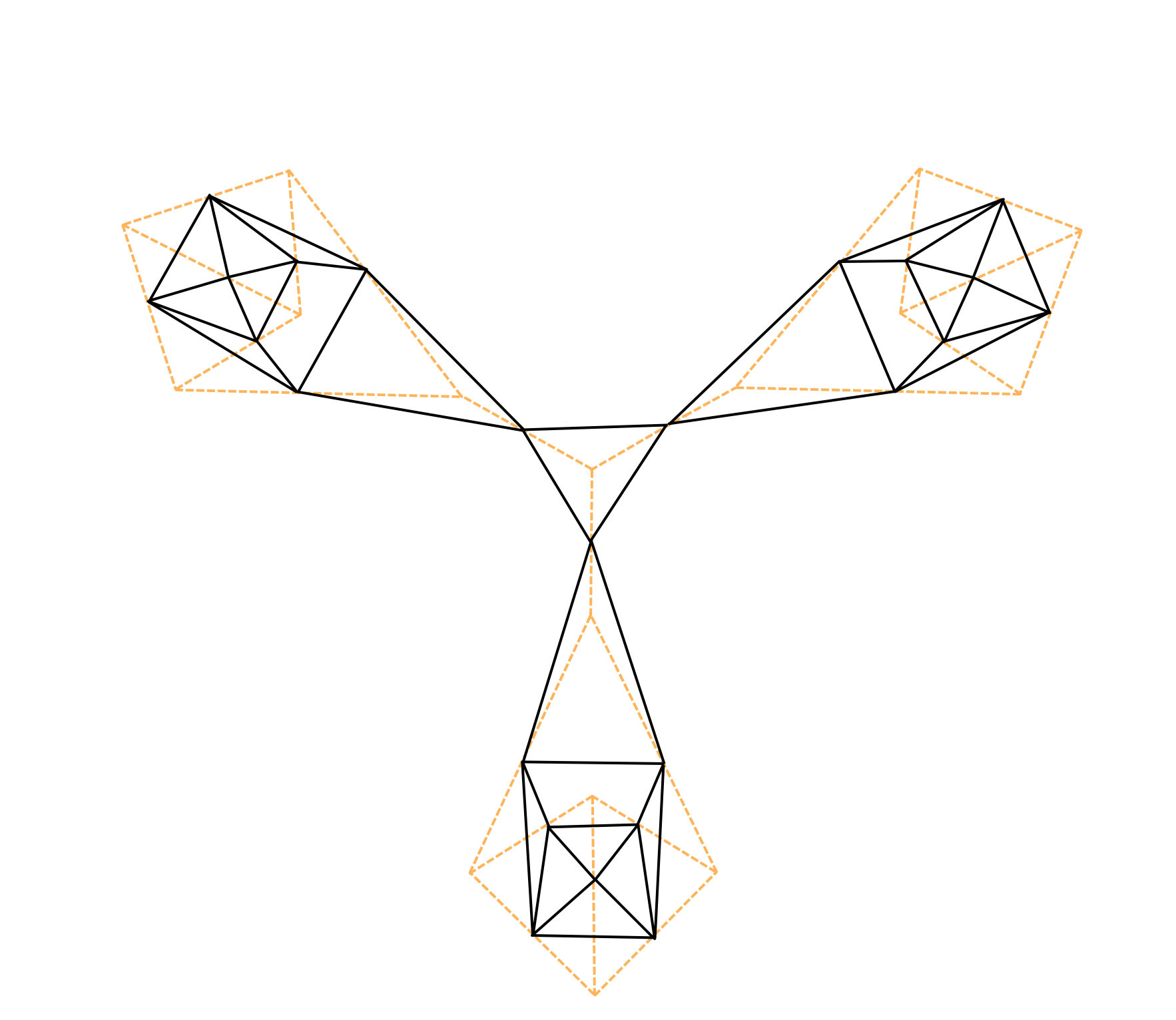}%
    \label{fig:counterexample_vor}
\end{subfigure}
\begin{subfigure}{}
    \includegraphics[clip,width=0.9\columnwidth]{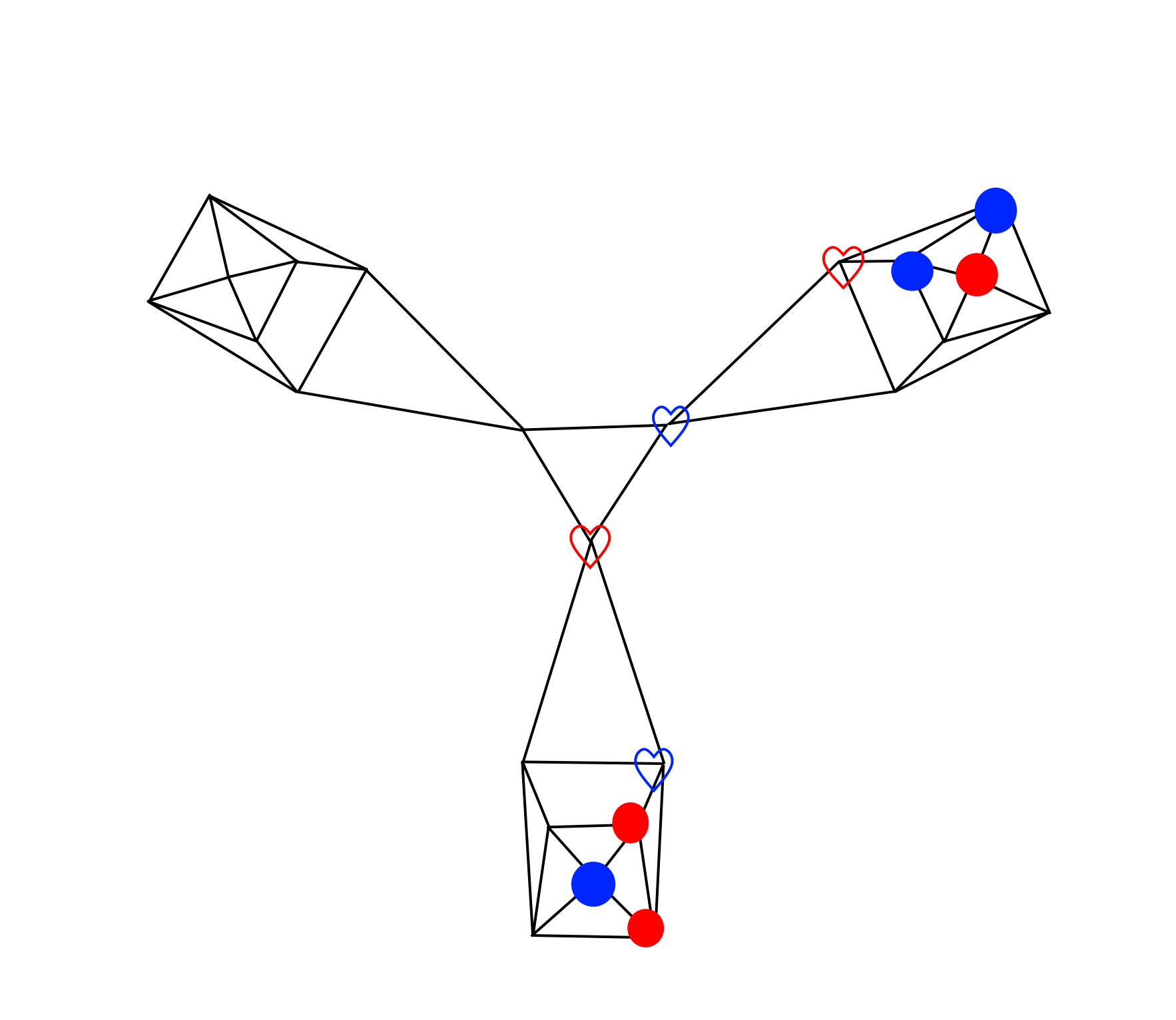}%
  \label{fig:counterexample_fbs}
\end{subfigure}

\caption{Top: Voronoi (dashed orange lines) diagram and kagomization (black) of a pinched graph that necessarily hosts Type-II flat band states. Bottom: Type-II flat band state on the kagomized graph. Blue (red) circles represent un-normalized state amplitude $+1$($-1$) and blue (red) hearts denote the Dirac string that connects the two disjoint secondary triangles, with amplitudes $+2$($-2$).}
\end{figure}

One might wonder if the Dirac strings are necessary and whether one might be able to use only non-intersecting even length loops that combine the odd secondary polygons. The graph we present in Fig. \ref{fig:counterexample_fbs} gives a counter example. There are 8 flat band states: 6 come from independent even length loops and 2 are these bridged Dirac string states.  Due to the pinched nature of the graph, it is impossible to connect the secondary triangles from the different petals with a loop, and one needs to use Type-II states.

\clearpage

\section{Nearest-level spacings distributions analyses}
\label{app:lsd}

The set of nearest level spacings ${\Delta E}$ for a system is defined as the absolute difference between all nearest neighbors in an ordered set of the system's eigenenergies. To define a distribution of the spacings, spacings are first normalized by the mean spacing to define $s = \Delta E / <\Delta E>$.  In the limit of infinite system size, nearest-level spacings distributions in Anderson model disordered systems have been shown to fit either a Poisson law, $P_P(s) = 1/\delta e^ {-(s/\delta)}$, or the Wigner surmise for the orthogonal class ($\beta = 1$), $P_O(s) = \frac{\pi}{2\delta} se^{-(\pi/4 s^2)}$ or a hybrid function, \citep{shklovskii1993}. These three cases correspond to systems with only localized states, only delocalized (metallic) states, and systems at the metal-insulator transition. 

\begin{figure}[htp]
\begin{subfigure}{}
    \includegraphics[clip,width=.8\columnwidth]{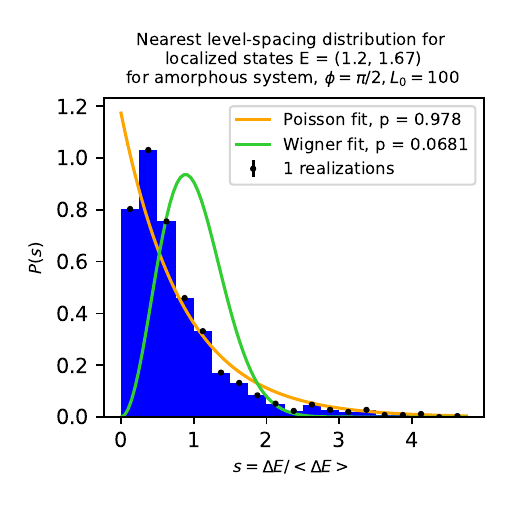}%
    \label{fig:D_loc_ess}
\end{subfigure}
\begin{subfigure}{}
    \includegraphics[clip,width=0.8\columnwidth]{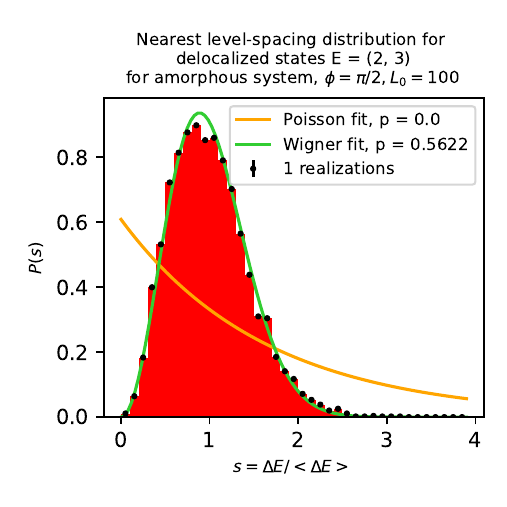}%
  \label{fig:D_deloc_ess}
\end{subfigure}
\caption{Top: Nearest level-spacings distributions of states in the localized energy range, averaged over 6 realizations of amorphous systems of size $L_0 = 100$. Error bars are in black for each bin. Best-fits for a Poisson law distribution and Wigner distribution are shown in orange and green, respectively. The distribution is likely to have come from the same distribution as the Poisson law. Bottom: Nearest level-spacings distributions of states in the delocalized energy range. The distribution is unlikely to have come from the same distribution as the Poisson law and likely to have come from the Wigner surmise.}
\end{figure}

\begin{figure}[htp]
\begin{subfigure}{}
    \includegraphics[clip,width=.8\columnwidth]{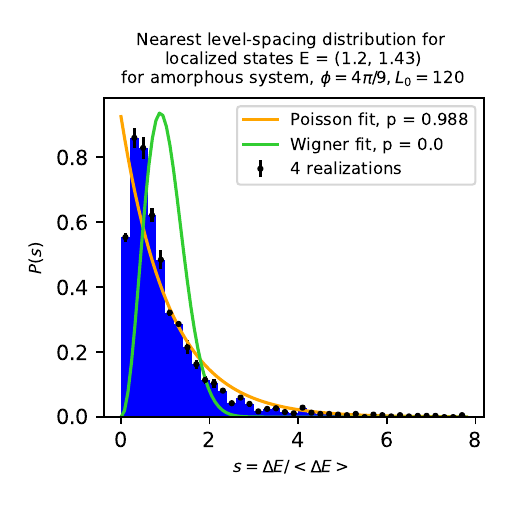}%
    \label{fig:A_loc_ess}
\end{subfigure}
\begin{subfigure}{}
    \includegraphics[clip,width=0.8\columnwidth]{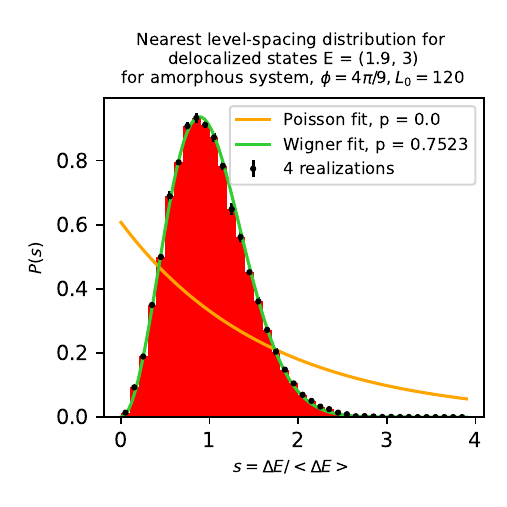}%
  \label{fig:A_deloc_ess}
\end{subfigure}
\caption{Top: Nearest level-spacings distributions of states in the localized energy range, averaged over 43 realizations of amorphous systems of size $L_0 = 120$ with $\phi = 4\pi/9$. Error bars are in black for each bin. Best-fits for a Poisson law distribution and Wigner distribution are shown in orange and green, respectively. The distribution was drawn from the same distribution as the Poisson law. Bottom: Nearest level-spacings distributions of states in the delocalized energy range. The distribution was drawn from the same distribution as the Wigner surmise.}
\end{figure}

\begin{figure}
    \centering
    \includegraphics[width = .8\columnwidth]{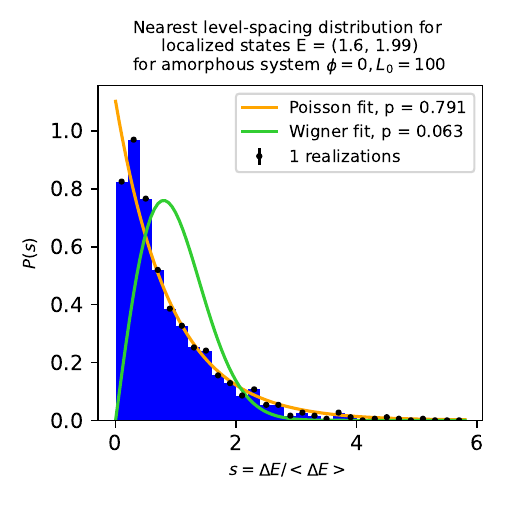}
    \caption{Nearest level-spacings distributions of states in the localized energy range for an amorphous systems of size $L_0 = 100$ with $\phi = 0$. The localized energy range is set to $E_{loc} = (1.6, 1.99)$, thus excluding the flat band at $E = 2$. Best-fits for a Poisson law distribution and Wigner distribution are shown in orange and green, respectively. The distribution was drawn from the same distribution as the Poisson law. } 
    \label{fig:B0_loc}
\end{figure}

An important difference from the Anderson model is that, for $\phi \neq 0$, systems break time reversal symmetry. Thus in our analysis we use the Wigner surmise for the unitary class (class A), $\beta = 2$, $P_W(s) = \frac{32}{\pi^2} s^2 e^{-4s^2/\pi}$ \citep{guhr-rmt}. The Wigner surmise for class D coincides with that of class A for states at sufficiently high energy \citep{zirnbauer-symclasses}. We define subsets of all eigenstates using cutoff $\tilde{\xi}/L < 1/3$ to approximate which states are localized, shown for example in Fig. \ref{fig:D_spectrum} in the main text. We examine the largest system size only. For each amorphous system of this size, we choose all energies within a range $R$. Then we select energies ${E_{i, R}}$ in this range and find the nearest level spacings ${\Delta E_{i, R}}$ and normalize this set by its mean. We then compute a histogram for the normalized spacings ${\Delta E_{i, R}}$. Using the same histogram bins, we repeat this procedure for all realizations, then find the average value and error bar for each bin. For each range $R$, we fit a Poisson distribution and a Wigner distribution and then compute the $p$-value from the Kolmogorov-Smirnoff statistic $p$ for each fit. A large value $p \sim 1$ indicates that the data is likely to have come from this type of distribution whereas $p \sim 0$ indicates that the data comes from a different distribution. Results of these analyses are shown in Fig. \ref{fig:D_loc_ess}, \ref{fig:A_loc_ess}, \ref{fig:B0_loc}

In amorphous systems with $\phi = 0$, in symmetry class AI, we find that states tend to localize near the upper tail of the single dispersive band. In Supplemental Figure \ref{fig:B0_loc}, we execute a nearest level-spacings analysis of the states suggested to localize, and confirm that these states are indeed localized. Note that since these systems are time-reversal-symmetric, we use the Wigner surmise for the orthogonal class ($\beta = 1$), $P_O(s) = \frac{\pi}{2\delta} s e^{-(\pi/4 s^2)}$. Our analysis suggests that states near the flat band at $E = 2$ are localized. An intriguing question for future work is to determine whether adjacency to the flat band plays a role in localizing these states.

\newpage
\bibliography{references}

\end{document}